\makeatletter
\declare@file@substitution{revtex4-1.cls}{revtex4-2.cls}
\makeatother
\documentclass[twocolumn, times, tighten,twocolappendix, a4paper]{aastex63}
\usepackage{amsmath,amssymb}
\usepackage[T1]{fontenc}
\usepackage{graphicx,multirow,hyperref,url,color,xspace}
\usepackage{lineno}

\usepackage{rotating}
\usepackage{subfigure}

\newcommand{\msun}{{M}_{\sun}}

\newcommand{\swift}{{\textit{Neil Gehrels Swift}}\xspace}
\newcommand{\suzaku}{{\textit{Suzaku}}\xspace}

\newcommand{\nustar}{{\textit{NuSTAR}}\xspace}
\newcommand{\nicer}{NICER\xspace}
\newcommand{\integral}{{\textit{INTEGRAL}}\xspace}

\newbox\grsign \setbox\grsign=\hbox{$>$} \newdimen\grdimen \grdimen=\ht\grsign
\newbox\simpropbox
\setbox\simpropbox=\hbox{\raise.5ex\hbox{$\propto$}\llap
 {\lower.5ex\hbox{$\sim$}}}\ht2=\grdimen\dp2=0pt

\begin{document}

\title{What Is the Black Hole Spin in Cyg X-1?}

\author[0000-0002-0333-2452]{Andrzej A. Zdziarski}
\affiliation{Nicolaus Copernicus Astronomical Center, Polish Academy of Sciences, Bartycka 18, PL-00-716 Warszawa, Poland; \href{mailto:aaz@camk.edu.pl}{aaz@camk.edu.pl}}

\author[0000-0003-3499-9273]{Swadesh Chand}
\affiliation{Inter-University Center for Astronomy and Astrophysics, Pune 411007, India}

\author[0000-0002-6051-6928]{Srimanta Banerjee}
\affiliation{Inter-University Center for Astronomy and Astrophysics, Pune 411007, India}

\author[0000-0001-7606-5925]{Micha{\l} Szanecki}
\affiliation{Faculty of Physics and Applied Informatics, {\L}{\'o}d{\'z} University, Pomorska 149/153, PL-90-236 {\L}{\'o}d{\'z}, Poland}

\author[0000-0002-1622-3036]{Agnieszka Janiuk}
\affiliation{Center for Theoretical Physics PAS, Al. Lotnikow 32/46, 02-668 Warsaw, Poland}

\author[0000-0001-6276-7045]{Piotr Lubi\'nski}
\affiliation{Institute of Physics, University of Zielona G\'{o}ra, Licealna 9, 
PL-65-417 Zielona G\'{o}ra, Poland}

\author[0000-0002-8541-8849]{Andrzej Nied{\'z}wiecki}
\affiliation{Faculty of Physics and Applied Informatics, {\L}{\'o}d{\'z} University, Pomorska 149/153, PL-90-236 {\L}{\'o}d{\'z}, Poland}

\author[0000-0003-1589-2075]{Gulab Dewangan}
\affiliation{Inter-University Center for Astronomy and Astrophysics, Pune 411007, India}

\author[0000-0002-7609-2779]{Ranjeev Misra}
\affiliation{Inter-University Center for Astronomy and Astrophysics, Pune 411007, India}

\begin{abstract}
We perform a detailed study of the black hole spin of Cyg X-1, using accurate broad-band X-ray data obtained in the soft spectral state by simultaneous NICER and NuSTAR observations, supplemented at high energies by INTEGRAL data. We use the relativistic disk model {\tt kerrbb} together with different models of the Comptonization high energy tail and the relativistically-broadened reflection features. Unlike most previous studies, we tie the spin parameters of the disk and relativistic broadening models, thus combining the continuum and reflection methods of spin determination. We also consider a likely increase of the disk color correction due to a partial support of the disk by large scale magnetic fields. We find that such models yield the spin parameter of $a_*= 0.87^{+0.04}_{-0.03}$ if the disk inclination is allowed to be free, with $i= 39\degr^{+1}_{-1}$. Assuming $i=27\fdg 5$, as determined by optical studies of the binary, worsens the fit, but leads to similar values of the spin, $a_*= 0.90^{+0.01}_{-0.01}$. In addition, we consider the presence of a warm Comptonization layer on top of the disk, motivated by successful modeling of soft X-ray excesses in other sources with such a model. This dramatically lowers the spin, to $a_*\lesssim 0.1$, consistent with the spin measurements from black-hole mergers. On the other hand, if the natal spin of Cyg X-1 was low but now $a_*\approx 0.9$, a period of effective super-critical accretion had to take place in the past. Such accretion could be facilitated by photon advection, as proposed for ultraluminous X-ray sources. 
\end{abstract}

\section{Introduction} \label{intro}

A highly important result obtained in recent years from gravitational wave studies is that the spins of merging binary black holes (BHs) are general low. The most recent population study of \citet{Abbott23} presents the results from 70 binary BH mergers. The effective spin parameter, which is an average of the individual spins weighted by the masses and the misalignment angles, has the mean value of $\approx 0.06$, and it was $<0.6$ in the sample. The distribution of the individual spins peaks at $a_*= 0.13^{+0.12}_{-0.11}$, and most of them are $\lesssim$0.4. Since accretion would spin up BHs (unless it is retrograde, which can happen in a minority of cases only), an implication of the results of \citet{Abbott23} is that most BHs are formed with low natal spins. This can happen if the stellar core during expansion (occurring when leaving the main sequence) remains strongly coupled to the outer envelope. This rules out the Geneva stellar models with moderate angular momentum transfer \citep{Ekstrom12}. On the other hand, the above results are consistent with the standard MESA stellar models with efficient transport \citep{Spruit02, Belczynski20}, predicting typically $a_*\sim 0.1$. 

Progenitors of binary BH mergers can be BH X-ray binaries (XRBs) with high-mass donors. We know three such systems, namely Cyg X-1, LMC X-1 and M33 X-7. Their published spin values are all high: $>$0.9985 \citep{Zhao21_CygX1}, $0.92^{+0.05}_{-0.07}$ \citep{Gou09}, and $0.84 \pm 0.05$ \citep{Liu08}, respectively. Those authors stress that those spins have to be natal, which, in turn, is in conflict with the result inferred from the observed spins of merging BHs. In the case of Cyg X-1, Eddington-limited accretion increases the BH mass on the time scale of $\approx 40$\,Myr (see Section \ref{spinup}), which is an order of magnitude longer than the lifetime of Cyg X-1 of about 4 Myr \citep{Miller-Jones21}. 

Thus, there is an acute disagreement between the low spins inferred for mergers and the high spins determined from modelling the electromagnetic emission of BH XRBs. The method used for the spin determination of the three BH XRBs with high-mass donors was the continuum method, based on fitting the shape of the disk continuum in the soft spectral state \citep{McClintock14}. The method assumes the disk extending to the innermost stable circular orbit (ISCO; as confirmed observationally, e.g., \citealt{GD04, Steiner11}), and that we know the BH mass and the distance. High-energy tails, commonly appearing beyond the disk spectra, are attributed to Comptonization of the disk photons in a corona above the disk. 

A further major assumption of the continuum model as used so far is that the disk is described by the standard accretion model \citep{NT73, Davis05, Davis06}. All of the three studies of the three high-mass XRBs were done using the {\tt kerrbb2} model \citep{McClintock06}. It uses the GR treatment of \citet{Li05} and the calculations of the local disk spectra of \citet{Davis05} and \citet{Davis06} for the viscosity parameter \citep{SS73} of $\alpha=0.1$.

However, as it is well known, the standard model fails to describe a lot of astrophysical phenomena. It predicts the disk to be viscously and thermally unstable when dominated by radiation pressure \citep{Lightman74, Shakura76}. To the contrary, observations of the X-ray emission in the soft state of BH XRBs show them to be very stable \citep{GD04}. Another problem involves the value of $\alpha$. Observationally, low-mass BH XRBs show $\alpha\simeq 0.2$--1 \citep{Tetarenko18_alpha}, while local simulations of the standard disk give $\alpha \sim 0.01$ (e.g., \citealt{Davis10, Simon11}). In the case of AGNs, the standard model predicts disk gravitational fragmentation, which is contrary to observations (e.g., \citealt{Begelman07}). Then, disk sizes inferred from microlensing are larger by a factor of a few than those predicted by standard disk theory, e.g., \citet{Chartas16}. Also, AGN variability time scales are often much shorter than those for standard disks (e.g., \citealt{Lawrence18}). 

Most of these problems can be resolved in the model of so-called magnetically elevated disks, in which the pressure of toroidal magnetic field dominates over other forms of the pressure \citep{Begelman07, Begelman17}, with a weak net vertical magnetic field also required. Those disks are stable in the radiative-pressure dominated regime, yield $\alpha\simeq 0.1$--1 \citep{Salvesen16a}, and allow faster variability \citep{Dexter19}. Importantly, they have lower gas densities at the effective photosphere, which enhances electron scattering and can lead to a harder spectrum with respect to a disc without magnetic pressure support. Then, the effective color correction, $f_{\rm col}\equiv T_{\rm color}/T_{\rm eff}$, can be $\gtrsim$1.7. As mentioned above, these effects were not included in the spin measurements for the three high-mass BH XRBs, as they used the standard disk model with $\alpha=0.1$ (which is approximately equivalent to $f_{\rm col}\lesssim 1.7$). This discrepancy (also for low-mass XRBs) was pointed out by \citet{Salvesen21}. However, they concentrated on comparison between the results of the continuum method with that based on relativistic broadening of reflection spectra. The latter method is often uncertain, in particular because of the uncertainty about the form of the incident continuum and the controversial assumption of the disk extending to the ISCO also in the hard spectral state. Here, we instead use the continuum and reflection methods jointly in the soft state, in which the disk does extend to the ISCO. 

Furthermore, warm coronae ($kT_{\rm e}\sim 1$\,keV, $\tau_{\rm T}\gg 1$) on top of optically thick accretion disks have been invoked to explain soft X-ray excesses found in many AGNs (e.g., \citealt{MBZ98, Petrucci20, Ursini20, Xiang22}). Analogous soft X-ray excesses were also found in some BH XRBs, in particular in the soft state of GRS 1915+105 (a low-mass BH XRB), which was fitted by hybrid Comptonization including a low-temperature, high optical depth, thermal component \citep{ZGP01}. A similar result was found for the very high state of GRO J1655--40 \citep{Kubota01}. A warm coronal component was found in the hard state of Cyg X-1 \citep{Basak17}. Therefore, we also consider models with a warm, optically-thick, thermal Comptonization in a layer on top of a standard disk, as done before in \citet{Belczynski23} and \citet{Zdziarski24a}. 

We study the above issues for the case of Cyg X-1. Its distance and BH mass are well known, see Section \ref{results}. We use highly accurate and simultaneous X-ray observations of this source in the soft spectral state by \nicer and \nustar, and a quasi-simultaneous \integral observation. In Section \ref{data}, we describe the data and their variability properties. In Section \ref{model}, we describe the models used to spectrally fit the data. Section \ref{results} presents our results, which are discussed in Section \ref{discussion}. Finally, we give our main conclusions.  

\section{The data} \label{data}

\begin{table}\centering
\caption{Simultaneous observations of Cyg X-1 in the soft state with \nicer and \nustar} 
\begin{tabular}{cc}
\hline
Date & Date [MJD]\\
\hline
2018-04-15 & 58223\\
2018-05-27 & 58265\\
2018-08-11 & 58341\\
{\bf 2019-11-13} & {\bf 58800}\\
2023-05-25 & 60089\\
2023-06-20 & 60115\\
\hline
\end{tabular}
\label{soft}
\tablecomments{The observation with the weakest high-energy tail, studied here, is marked in bold. } 
\end{table}

\begin{table*}\centering
\caption{The log of the observations with the \nicer, \nustar and \integral} 
\begin{tabular}{cccccc}
\hline
Detector & Obs. ID & Start time [MJD] & End time [MJD] & Exposure [s] & Phase\\
\hline
NICER & 2636010102 & 58800.43889 & 58800.90162 & 13327 & 0.63--0.71\\
NuSTAR A & 80502335006 &58800.42093 & 58800.88622 & 11914 & 0.63--0.71\\
NuSTAR B & -- & 58800.42093 & 58800.88622 & 12246 & 0.63--0.71\\
INTEGRAL/ISGRI & 1620024 &58802.27764 & 58803.91486 & 78316& 0.96--0.25\\
\hline
\end{tabular}
\label{log}
\tablecomments{The orbital phases are according to the ephemeris of \citet{Gies08} (the null phase is at the superior conjunction of the BH).}
\end{table*}

\begin{figure*}
\centerline{\includegraphics[width=11cm]{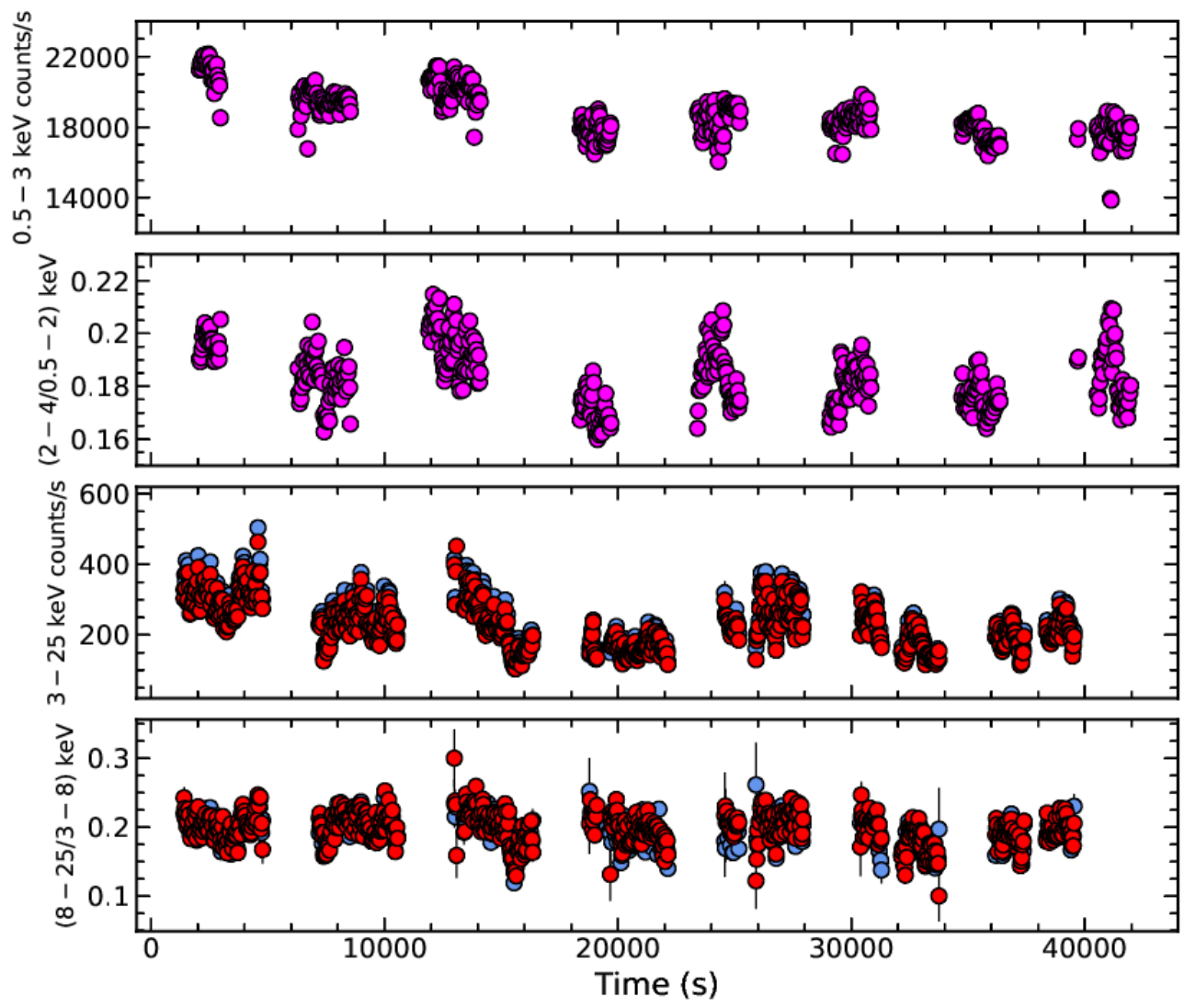}}
  \caption{The count rates and hardness ratios of NICER and \nustar. The top two panels show the NICER 0.5--3.0\,keV count-rate and the (2--4\,keV)/(0.5--2.0\,keV) count rate ratio. The bottom two panels show the \nustar 3--25\,keV count-rate and the (8--25\,keV)/(3--8\,keV) count rate ratio. The blue and red symbols correspond to the \nustar A and B units, respectively. The zero time corresponds to MJD 58800.42093. 
}\label{lc}
\end{figure*}

\begin{figure}
\centerline{\includegraphics[width=7.5cm]{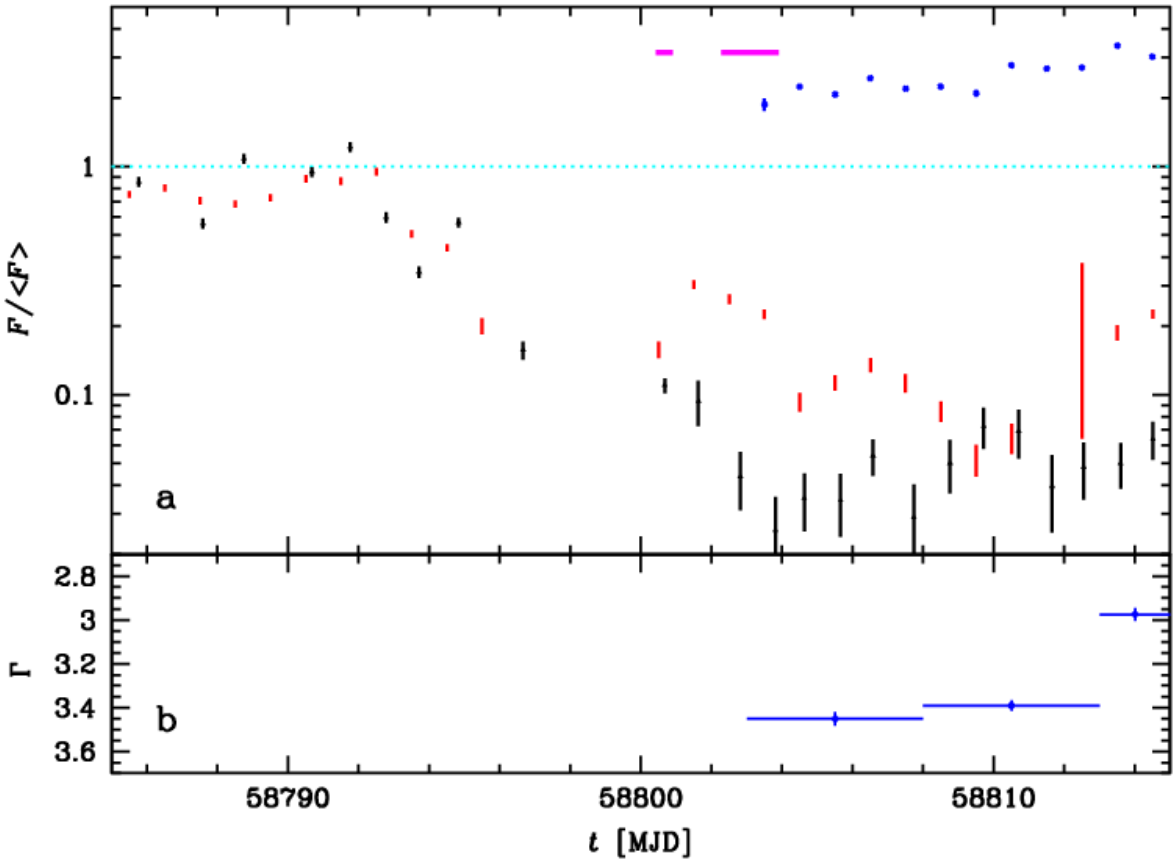}}
  \caption{(a) Monitoring daily light curves around the studied observation normalized to the hard-state averages (dotted line). The magenta bars show the durations of the \nicer/\nustar and \integral observations. The blue, red and black symbols show the fluxes from MAXI (2--20\,keV, $\langle F\rangle=1.0$\,cm$^{-2}$\,s$^{-1}$), BAT (15--50\,keV, $\langle F\rangle=0.17$\,cm$^{-2}$\,s$^{-1}$) and AMI-LA (15 GHz, $\langle F\rangle=12.7$\,mJy). (b) The 2--10\,keV photon spectral index estimated from the MAXI count rates. 
}\label{monitoring}
\end{figure}

The \nicer and \nustar instruments are uniquely suited for broad-band measurements of Cyg X-1, combining high effective areas with no significant pileup. We have selected simultaneous observations of Cyg X-1 in the soft state by both detectors. We have found 6 data sets, listed in Table \ref{soft}. Given our goal of measuring the BH spin, we have chosen the one with the weakest high energy tail beyond the disk spectrum, which occurred on 2019-11-13. In order to extend the high-energy coverage of the spectrum, we also included a contemporaneous observation by \integral, for which we used the ISGRI data. They fall inside the region of the extreme soft state according to the observed hard X-ray flux and photon index \citep{Lubinski20}. The log of the data is given in Table \ref{log}, and the count rates and hardness ratios for \nicer and \nustar are shown in Figure \ref{lc}. We see only moderate changes of the count rates and the hardness ratios. 

The \integral observation started about 1.4\,d after the end of the \nicer and \nustar ones. In order to assess the level of variability of the emission during that period, Figure \ref{monitoring} shows the public monitoring data from the MAXI (2--20\,keV; \citealt{Matsuoka09}), the \swift BAT (15--50\,keV; \citealt{BAT}), and the 15 GHz flux from the Arcminute Microkelvin Imager Large Array (AMI-LA) (\citealt{Hickish18}; as published in \citealt{Zdziarski23b}). We see that our observations were clearly in the soft state, and the variability of the 15--50\,keV flux during them was modest. Indeed, we find that the spectrum from the ISGRI agrees well in the slope with that from \nustar in the overlapping range, with the only difference in the normalization of the ISGRI spectrum being about a factor of $\approx$1.4 higher than that from \nustar. 

The \nicer, \nustar, and \integral data used for the spectral analysis are in the 0.3--10.0, 4--79 and 30--220\,keV ranges, respectively. We used the \nustar data only above 4\,keV because below it there were substantial systematic differences between the \nustar A and B modules and between the \nustar and \nicer data. We extracted the \nustar data as required for bright sources\footnote{We used the option {\tt statusexpr=\\"(STATUS==b0000xxx00xxxx000)\&\&(SHIELD==0)"}.}. During the standard data reduction of NICER, the script {\tt nicerl3-spect} added 1.5\% systematic error to each channel in the 0.34--9.0\,keV range, and somewhat more (up to 2.5\%) beyond that\footnote{\url{https://heasarc.gsfc.nasa.gov/lheasoft/ftools/headas/nicerl3-spect.html}}. We have not added any systematic errors to either the \nustar or \integral data. The \nicer and \nustar spectral data have been optimally binned \citep{Kaastra16} with the additional requirement of at least 20 counts per bin.

\section{The variability} \label{variability}

\begin{figure}
\centerline{\includegraphics[width=8cm]{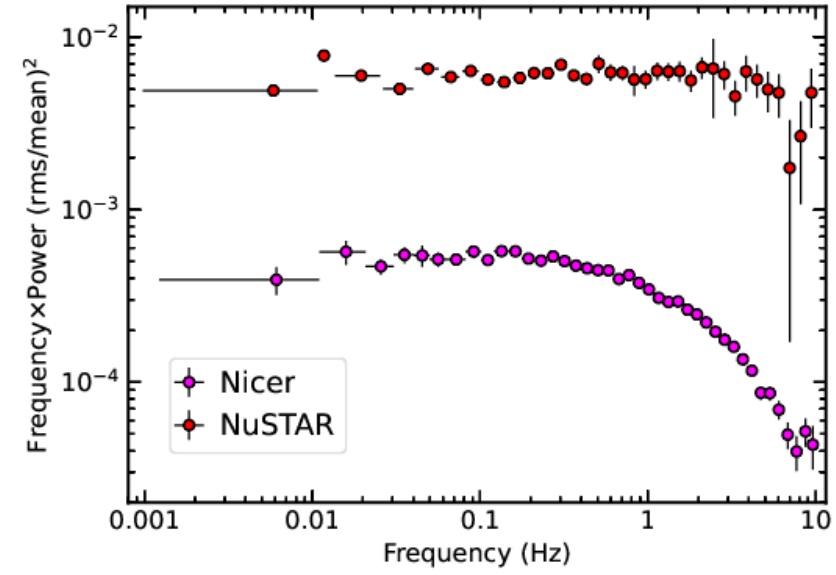}}
  \caption{The power spectra from NICER, 0.34--2.0 keV (magenta circles), and the co-spectra from \nustar, 5--30 keV (red circles). 
}\label{power}
\end{figure}

The source flux shows relatively low variability on short time scales. We have calculated the power spectra in the 0.34--2.0 (\nicer) and 5--30\,keV (\nustar) energy ranges with the white noise subtracted, see Figure \ref{power}. The latter is the co-spectrum calculated using the {\sc hendrics} package \citep{Bachetti15, Lazar21}. We see that the fractional variability squared per unit $\ln f$ is very low in the disk-dominated 0.34--2.0\,keV range, $\sim\! 5\times 10^{-4}$ at at $f\lesssim 0.3$\,Hz and fast declining at higher $f$. The decline shape appears to be well reproduced by the model of a linear damped harmonic oscillator \citep{Misra08}. At the corona-dominated 5--30\,keV range, the variability is larger, $\sim\! 6\times 10^{-3}$, without a clear high-$f$ cutoff. These results show that the coronal variability is not directly linked to the disk variability, and does not have a sharp cutoff at high frequencies, presumably corresponding to lower radii of the disk. We have computed the rms in the 0.01--1\,Hz range, as $4.88\pm 0.08\%$ and $16.91\pm 0.03\%$ in the energy ranges of 0.34--2 and 5--30\,keV, respectively. 

\begin{figure}
\centerline{\includegraphics[width=7.5cm]{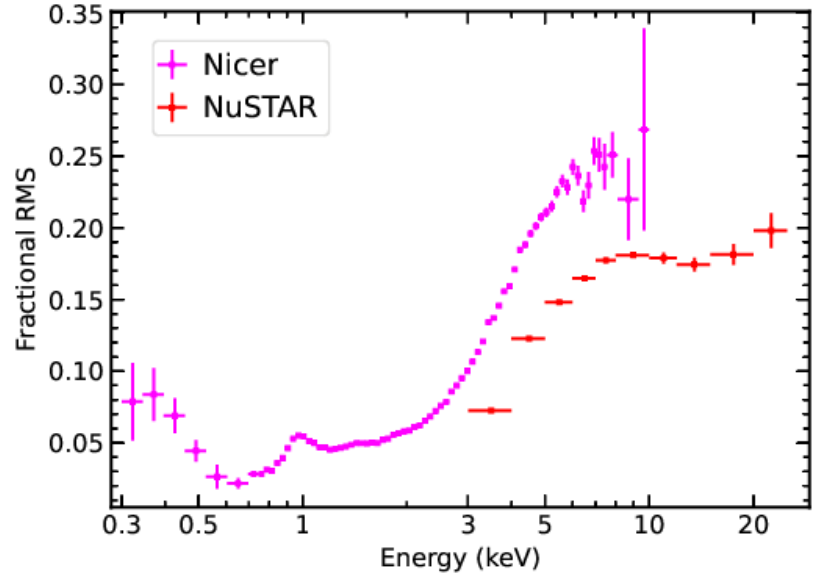}}
  \caption{The rms vs.\ photon energy for the frequency range of 0.01--1\,Hz. The lower values for \nustar (red) than for \nicer (magenta) in the overlapping energy range are due to the dead time of \nustar.
}\label{rms}
\end{figure}

We have then calculated the fractional rms within 0.01--1\,Hz as a function of the energy, see Figure \ref{rms}. We see very weak variability at low energies dominated by the accretion disk, and then the rms increases to much higher levels in the range dominated by the high-energy tail. Thus, we see a stable disk and a (moderately) variable corona, as first found for the soft state of Cyg X-1 by \citet{CGR01}. Notably, the \nicer data show significantly higher variability than those from \nustar. This effect appears to due to the presence of some dead time in the \nustar data, which artificially lowers their rms (\citealt{Bachetti18}; M. Bachetti, private communication). We have tested whether this effect could be due to the slightly different live times of the two instruments, shown in Figure \ref{lc}. For that, we have calculated the rms values based on the good-time intervals identical for both \nicer and \nustar, and found the dependencies almost identical to those shown in Figure \ref{rms}. Thus, the \nustar rms needs to be corrected upward, by $\Delta{\rm rms}\sim 0.05$. 

The \nicer rms increases from its minimum of $\approx$0.02 at 0.6\,keV to $\approx$0.25 at 6\,keV. This is naturally explained by both the coronal contribution increasing with the increasing energy and the disk radii dominating the emission increasing with $E$. Then, the rms has a pronounced dip with the depth of $\approx$0.03 at $\approx$6.5\,keV, which appears to be explained by the contribution of the narrow Fe K$\alpha$ due to remote scattering, present in the spectrum (Section \ref{results}). However, the rms dependence shows some peculiar features at low energies. First, there is a pronounced maximum at $\approx$1\,keV, which does not seem to correspond to any spectral feature, see Figure \ref{models} below. While we see there some residuals around 1\,keV, they are not stronger than others present at $E\lesssim 2$\,keV. We have searched for such a feature in the rms of a \nicer Crab observation, and found none, while we have found a similar feature in the \nicer observation of Cyg X-1 in the soft state on 2023-05-25. The rms increase is then unlikely to be of instrumental origin. It can be due to absorption and subsequent re-emission by the stellar wind of the donor in the Ne {\sc ix} and {\sc x} lines at $\approx$0.9 and 1\,keV, respectively. Then, the rms increases at $\lesssim$0.6\,keV, which is also likely to be due to reprocessing in the stellar wind (see \citealt{Lai22} for a study of the effect of the wind on the timing features of Cyg X-1).  

We searched for the presence of dips in the light curves. These events are sharp decreases of the soft X-ray count rates and associated increases of the hardness, caused by absorption in clumps of the stellar wind. Such events occur in Cyg X-1 preferentially around the orbital phase 0 (the superior conjunction), when the line of sight to the observer is the longest (e.g., \citealt{PZI08}). We have not detected any, which is consistent with the \nicer observation being away from the superior conjunction, see Table \ref{log}. 

\section{The models} \label{model}

We have accurate determination of the binary parameters; the binary inclination is $i_{\rm b}=27\fdg 5_{-0.6}^{+0.8}$, the mass of the BH is $M=21.2\pm 2.2 \msun$ and the distance to the source is $D= 2.22^{+0.18}_{-0.17}$\,kpc (given here as the median values with 68\% uncertainties; \citealt{Miller-Jones21}). Hereafter, we assume the above median values except that the inclinations of the disk and the reflection component (see below) are either kept at $i=i_{\rm b}$ or allowed to be free.

We fit the spectra using the {\sc{xspec}} environment \citep{Arnaud96}, with the uncertainties calculated for 90\% confidence ($\Delta\chi^2 \approx 2.71$; \citealt{Lampton76}). Differences between the \nicer and \nustar calibrations are treated using the {\tt plabs} model by multiplying the model spectra by $K E^{-\Delta\Gamma}$. We assume $K=1$ and $\Delta\Gamma=0$ for the \nustar A detector, as well as $\Delta\Gamma= 0$ for \nustar B \citep{Madsen22} and \integral (necessary due to the relatively large statistical uncertainties). We set the photon energy grid (as necessary for convolution models) by the {\sc xspec} command {\tt energies 0.001 2000 2000 log}. 

The Galactic column density in the direction of Cyg X-1 is $\approx 7.1\times 10^{21}$\,cm$^{-2}$ \citep{HI4PI}, and we find similar values from spectral fitting. We assume the abundances of \citet{Wilms00}, as given in the model {\tt tbabs} except for those for O and Fe, which we fit within the ranges of a factor of 2. To account for absorption by the stellar wind, we use a table model of an ionized absorber based on the {\tt XSTAR} code (\citealt{xstar}) developed by \citet{Tomsick14}, which we denote here as {\tt wind\_abs}. It has the solar abundances, the H number density of $10^{12}$\,cm$^{-3}$ and the turbulent velocity of $300$\,km\,s$^{-1}$. Its parameters are the ionization parameter, $\xi_{\rm w}$, and the column density, $N_{\rm Hw}$, defined in the ranges of $10^2 \leq \xi_{\rm w}\leq 10^5$\,erg\,cm\,s$^{-1}$ and $1\times 10^{21} \leq N_{\rm Hw} \leq 5\times 10^{22}$\,cm$^{-2}$, respectively.

We use the {\tt kerrbb} model \citep{Li05} to model the accretion disk. The local emission is assumed to be a blackbody with a color correction, $f_{\rm col}$ (see discussion in Section \ref{intro}). The parameters are $f_{\rm col}$, the spin parameter, $a_*$, and the mass accretion rate, $\dot M_{\rm disk}$. The disk inner radius is at the ISCO, at which the zero stress boundary condition is assumed. We include the disk self-irradiation.

Cyg X-1 in the soft state shows relatively strong high-energy tails \citep{G99, McConnell02, Tomsick14, Walton16}. They are most likely produced by Comptonization by electrons in a hot corona above the disk. We use three different convolution codes to account for it. In all of them, we assume the seed photons are from the disk emission, and a fraction, $f_{\rm cov}$, of the disk photons enter the corona. The simplest one is {\tt simplcut} \citep{Steiner09}, which phenomenologically moves photons from a seed distribution into a power law with an exponential cutoff. We then use two physical Comptonization codes allowing for the electron distribution to be hybrid, i.e., Maxwellian with a high-energy tail. 

The first is {\tt eqpair} \citep{PC98,Coppi99,G99}, and we use its convolution version, {\tt ceqpair}\footnote{\url{https://github.com/mitsza/eqpair_conv}}. This model treats Compton scattering, e$^\pm$ pair production/annihilation, bremsstrahlung, and the energy exchange between thermal and non-thermal parts of the electron distribution by Coulomb scattering. Some of electrons are accelerated, and the thermal ones are heated. The resulting non-thermal part of the distribution and the electron temperature, $kT_{\rm e}$, are calculated self-consistently. The thermal part is also described by the Thomson optical depth, $\tau_{\rm T}$. Some of the parameters are defined by the compactness,
\begin{equation}
\ell\equiv \frac{L\sigma_{\rm T}}{R m_{\rm e} c^3},
\label{ell}
\end{equation}
where $L$ is a power supplied to the source, $R$ is the characteristic size, and $\sigma_{\rm T}$ is the Thomson cross section. Apart from entering the definition of compactness, the size affects only bremsstrahlung (negligible for Cyg X-1) and the Coulomb logarithm. We define the hard compactness, $\ell_{\rm h}$, giving the power supplied to the electrons, and the soft one, $\ell_{\rm s}$, corresponding to the power in the seed photons. For the disk/corona geometry, the latter corresponds to one side of the disk. We also define the acceleration compactness, $\ell_{\rm nth}$, and one for the heating of the thermal electrons, $\ell_{\rm th}$ ($\ell_{\rm h}=\ell_{\rm nth}+\ell_{\rm th}$). The ratio $\ell_{\rm nth}/\ell_{\rm h}$ is a parameter. The rate at which non-thermal electrons are injected is a power law, $\dot N_+(\gamma) \propto \gamma^{-\Gamma_{\rm inj}}$, between $\gamma_{\rm min}$ and $\gamma_{\rm max}$ (while the steady state distribution is different from a power law). 

The other is {\tt comppsc}\footnote{\url{https://github.com/mitsza/compps_conv}}, which is a convolution version of the {\tt compps} iterative-scattering code \citep{PS96}. This model approximates the high-energy tail as a power law in the electron momentum, parametrized by its index, $p$, and the minimum and maximum Lorentz factors, $\gamma_{\rm min}$ and $\gamma_{\rm max}$, respectively. The Maxwellian and the power law intersect at $\gamma_{\rm min}$. The other parameters are $kT_{\rm e}$ and $\tau_{\rm T}$. We assume a slab geometry.

As discussed in Section \ref{intro}, we also take into account possible dissipation in surface layers of the disk. We model that by fully covering the disk by a warm scattering layer with $kT_{\rm e}\sim 1$\,keV and $\tau_{\rm T}\gg 1$. For that, we use the {\tt thcomp} code \citep{Z20_thcomp} with free $kT_{\rm e}$ and $\tau_{\rm T}$.  

The spectra also show relativistic reflection/reprocessing features, e.g., \citet{Tomsick14}. There are two major problems in treating it in soft states of BH XRBs; a high disk temperature, causing strong collisional ionization, and a relatively high density of the reflector. The former is not treated in available codes, and the latter is treated in some cases, but not for the emission of a hybrid electron distribution, such as present in Cyg X-1. Furthermore, the spectral shape of the irradiating radiation is best described by hybrid Comptonization, which is different from either thermal Comptonization or an e-folded power law. We thus use the convolution model, {\tt xilconv}. It is based on the {\tt relxill} opacity tables of \citet{Garcia13} and the Green's functions of \citet{MZ95}. The reflection features are then relativistically broadened, which we treat by using the convolution model {\tt relconv} \citep{Dauser10}. That model assumes a power-law radial distribution of the irradiation, $\propto R^{-\beta}$, and we allow for $\beta$ to be a free parameter. We assume the inner radius of the reflection emission to be at the ISCO.

\section{Results}
\label{results}

\begin{figure}
\centerline{\includegraphics[width=6.cm]{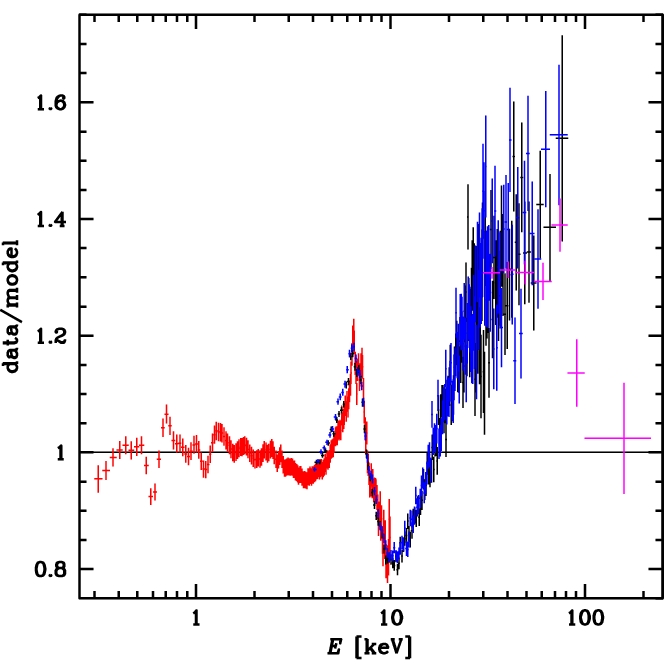}}
  \caption{The residuals to the fit with absorbed relativistic disk and a Comptonization tail (modeled by {\tt simpl}) to the \nicer (red), \nustar (black and blue) and \integral (magenta) data. Canonical reflection features are seen. 
}\label{profile}
\end{figure}

\begin{table*}
\caption{The results of spectral fitting
}
   \centering\begin{tabular}{lccccc}
\hline
Component & Parameter & Model 1 & Model 2 & Model 3 & Model 4\\
\hline
Absorption & $N_{\rm H}$ $[10^{21}$\,cm$^{-2}$] & $7.23^{+0.01}_{-0.01}$ & $7.20^{+0.04}_{-0.03}$ & $7.24^{+0.04}_{-0.04}$  & $7.53^{+0.19}_{-0.09}$  \\
& $Z_{\rm O}$ & $1.10^{+0.02}_{-0.01}$ & $1.09^{+0.02}_{-0.02}$                                   & $1.10^{+0.02}_{-0.02}$  & $1.10^{+0.02}_{-0.02}$  \\
& $Z_{\rm Fe}$ &$0.50^{+0.02}$ & $0.50^{+0.02}$& $0.50^{+0.05}$ & $0.50^{+0.05}$ \\                                                          
& $N_{\rm H,wind}$ $[10^{21}$\,cm$^{-2}$] & $2.0^{+0.9}_{-0.3}$ & $3.6^{+0.4}_{-0.7}$             & $3.7^{+0.4}_{-0.6}$     & $3.5^{+0.3}_{-0.8}$     \\
& $\log_{10}(\xi_{\rm wind}$) & $3.67^{+0.04}_{-0.08}$ & $3.80^{+0.05}_{-0.07}$                   & $3.77^{+0.06}_{-0.05}$  & $3.77^{+0.06}_{-0.05}$  \\
\hline                                                                                                                      
Disk & $a_*$ & $0.87^{+0.04}_{-0.03}$ & $0.90^{+0.02}_{-0.01}$                                    & $0.986^{+0.002}_{-0.001}$  & $0.00^{+0.07}$  \\
& $i\,[\degr]$ & $39^{+1}_{-1}$ & 27.5f & 27.5f  & $37^{+1}_{-1}$  \\                                                                
& $f_{\rm col}$ & $1.74^{+0.07}_{-0.09}$ & $1.95^{+0.01}_{-0.03}$                                 & 1.7f                    & $1.26^{+0.17}_{-0.08}$  \\
& $\dot M_{\rm disk}\, [10^{18}{\rm g\,s}^{-1}]$ & $0.34^{+0.01}_{-0.02}$ & $0.30^{+0.01}_{-0.01}$& $0.23^{+0.01}_{-0.01}$  & $0.42^{+0.01}_{-0.02}$  \\
\hline                                                                                                                      
Coronal  & $\ell_{\rm h}/\ell_{\rm s}$ & $0.31^{+0.04}_{-0.04}$ & $0.44^{+0.01}_{-0.03}$          & $0.47^{+0.03}_{-0.02}$  & $0.66^{+0.14}_{-0.04}$  \\
 Comptonization & $\ell_{\rm nth}/\ell_{\rm h}$  & $0.61^{+0.01}_{-0.05}$ &  $0.62^{+0.01}_{-0.01}$     & $0.59^{+0.01}_{-0.01}$  & $0.93^{+0.01}_{-0.04}$  \\
& $\tau_{\rm T}$ & $0.18^{+0.02}_{-0.05}$ & $0.34^{+0.01}_{-0.05}$                                & $0.38^{+0.01}_{-0.04}$  & $0.79^{+0.07}_{-0.06}$  \\
& $\Gamma_{\rm inj}$ & $0.2^{+0.3}_{-0.2}$ & $0^{+0.4}$                                           & $0^{+0.1}$     & $0.6^{+0.3}_{-0.3}$     \\
& $\gamma_{\rm max}$ & $6.5^{+0.3}_{-0.6}$ & $6.5^{+0.2}_{-0.6}$ &                                $6.5^{+0.2}_{-0.8}$       &$5.9^{+0.2}_{-0.4}$      \\
&$f_{\rm cov}$ & $0.18^{+0.03}_{-0.04}$ &  $0.10^{+0.01}_{-0.03}$ &                               $0.09^{+0.01}_{-0.01}$    &$0.07^{+0.01}_{-0.01}$   \\
 & $kT_{\rm e}$ [keV] &  62& 51 & 48 & 32  \\                                                                                    
\hline                                                                                                                      
Disk skin & $\tau_{\rm T}$ & -- & -- &                                                            --                        &$32^{+1}_{-1}$           \\
Comptonization & $kT_{\rm e}$ [keV] & -- & -- &                                                   --                        &$0.43^{+0.01}_{-0.01}$   \\
\hline                                                                                                                      
Reflection and  & ${\cal R}$ & $0.9^{+0.1}_{-0.1}$ & $1.2^{+0.1}_{-0.1}$                          &$1.3^{+0.1}_{-0.1}$      &$1.2^{+0.1}_{-0.1}$      \\
narrow line   &$\log_{10}\xi$ & $3.95^{+0.08}_{-0.08}$ & $4.28^{+0.02}_{-0.04}$ &                 $4.30^{+0.01}_{-0.11}$    &$4.10^{+0.01}_{-0.10}$   \\
& $\beta$ & $3.1^{+0.2}_{-0.2}$ &  $2.2^{+0.1}_{-0.1}$ &                                          $2.2^{+0.1}_{-0.1}$       &$5.2^{+1.2}_{-0.5}$      \\
& $Z_{\rm Fe}$ & $5.2^{+0.8}_{-0.5}$ &  $6.0_{-0.3}$                                              & $6.0_{-0.2}$            & $6.0_{-0.6}$            \\
&$E_{\rm line}$ [keV] &$6.40^{+0.04}$ & $6.44^{+0.02}_{-0.04}$ &                          $6.43^{+0.02}_{-0.03}$    &$6.43^{+0.01}_{-0.03}$   \\
& $N_{\rm line}\,[10^{-4}\,{\rm cm}^{-2}{\rm s}^{-1}]$ & $8.4^{+1.5}_{-1.5}$& $6.5^{+1.5}_{-1.3}$ &$7.0^{+1.5}_{-1.4}$    &$8.6^{+1.4}_{-1.3}$\\ 
\hline                                                                                                                      
Cross-calibration&$\Delta\Gamma_{\rm NICER}$&$-0.055^{+0.011}_{-0.011}$& $-0.047^{+0.007}_{-0.012}$   &$-0.053^{+0.006}_{-0.006}$ &$-0.056^{+0.007}_{-0.011}$ \\
& $K_{\rm NICER}$ & $1.03^{+0.02}_{-0.02}$ & $1.04^{+0.01}_{-0.02}$                               & $1.03^{+0.01}_{-0.01}$  & $1.03^{+0.01}_{-0.02}$  \\
&$K_{\rm NuSTAR B}$ & $0.992^{+0.002}_{-0.002}$ & $0.992^{+0.002}_{-0.002}$                       &$0.992^{+0.002}_{-0.002}$&$0.992^{+0.002}_{-0.001}$\\
& $K_{\rm INTEGRAL}$ & $1.37^{+0.02}_{-0.01}$ & $1.36^{+0.01}_{-0.01}$                            & $1.36^{+0.02}_{-0.02}$  & $1.37^{+0.01}_{-0.02}$  \\
\hline                                                                                                                      
& $\chi_\nu^2$  & 741/607 & 760/608                                                               & 765/609                 & 705/605                 \\
\hline
\end{tabular}
\tablecomments{Models 1, 2 and 3 = {\tt plabs*tbfeo*wind\_abs(relconv*xilconv*ceqpair*kerrbb + ceqpair*kerrbb + kerrbb+gaussian)}; Model 4 = {\tt plabs*tbfeo*wind\_abs(relconv*xilconv*ceqpair*thcomp*kerrbb + ceqpair*thcomp*kerrbb + thcomp*kerrbb+gaussian)}. The four additive components represent the reflection of the scattered radiation, the scattered emission going to the observer, the unscattered disk emission and the narrow line from distant fluorescence. In coronal Comptonization, $kT_{\rm e}$ is the best fit value (not a free parameter). $\xi\equiv 4{\pi}F_{\rm{irr}}/n$ is the ionization parameter, where $F_{\rm{irr}}$ is the irradiating flux. We have constrained $a_*\geq 0$, $0.5\leq Z_{\rm Fe}\leq 6$, $\Gamma_{\rm inj}\geq 0$, and 6.4\,keV\,$\leq E_{\rm Fe}\leq 7.0$\,keV.}
\label{fits}
\end{table*}

\subsection{Phenomenological models}
\label{powerlaw}

We first fit the data with a model consisting of {\tt kerrbb}, which provides seed photons for the high-energy tail, modeled by {\tt simpl}. The spectrum undergoes both the Galactic and wind absorption, but no reflection features are included. We show the residuals to the fit in Figure \ref{profile}. We see they are typical to reflection, consisting of a moderately broad Fe K line, an Fe K edge, and a reflection hump above about 20\,keV.

\begin{figure*}
\centerline{\includegraphics[width=7.2cm]{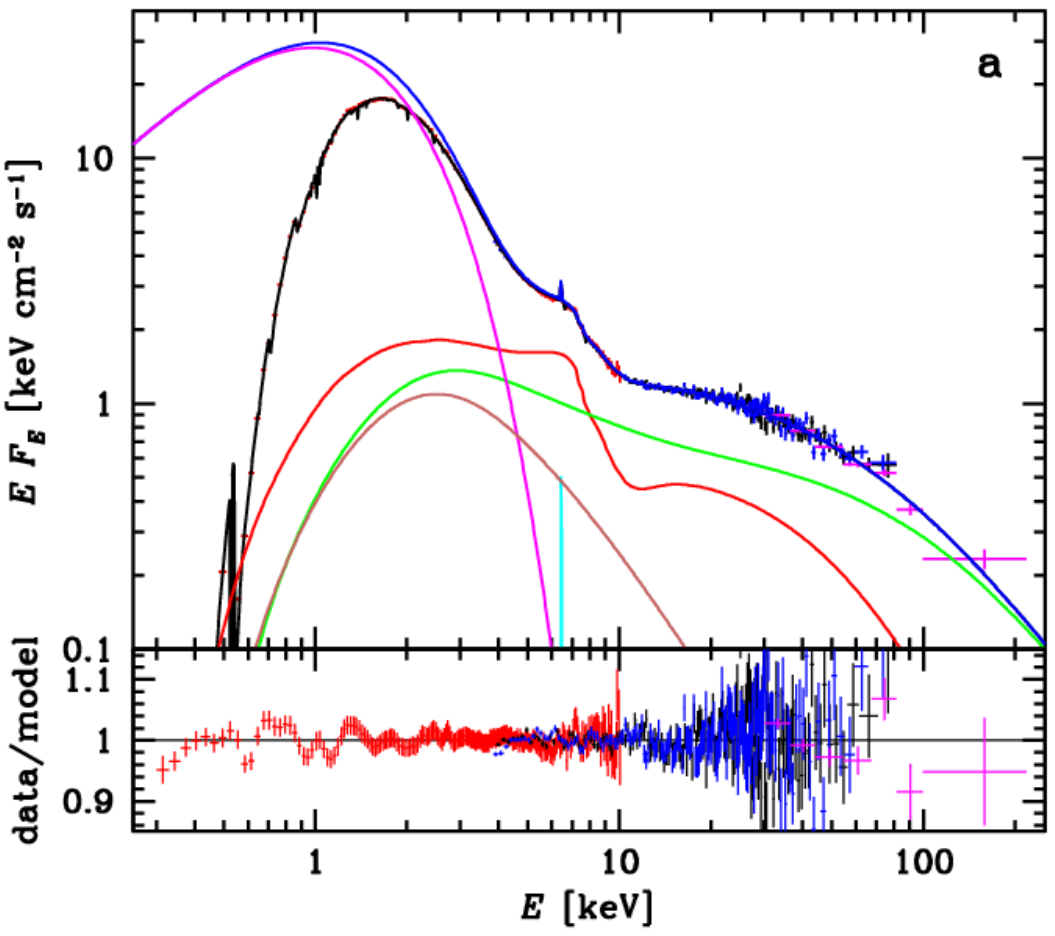}\includegraphics[width=7.2cm]{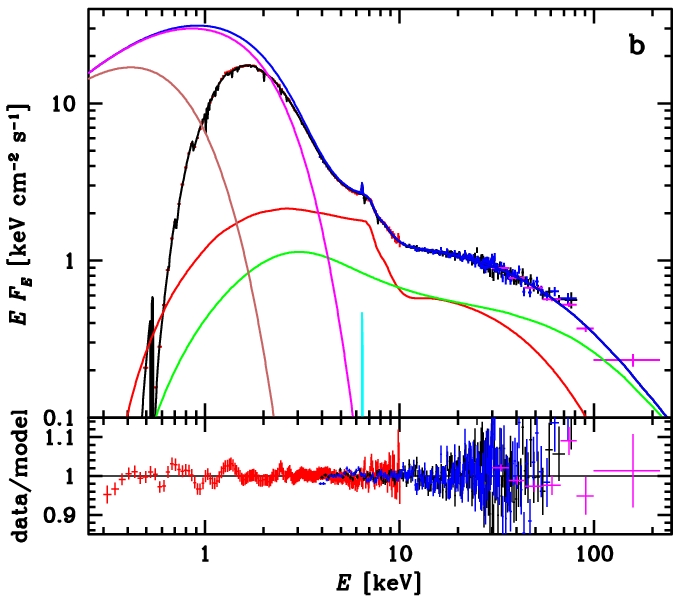}}
\caption{The \nicer (red), \nustar (black and blue) and \integral (magenta) unfolded spectra (top panel) and data-to-model ratios (bottom panel) for (a) Model 1 and (b) Model 4. The spectra are normalized to \nustar A. The total model spectrum and the unabsorbed one are shown by the solid black and blue curves, respectively. The unabsorbed disk, scattered, reflected, and narrow Fe K$\alpha$ components are shown by the magenta, green, red and cyan curves, respectively. In (a), the brown curve shows the disk spectrum scattered by the thermal electron component of the hybrid distribution, and in (b), it shows the underlying disk spectrum before going through the Comptonization top layer. 
}\label{models}
\end{figure*}

\begin{figure*}
\centerline{\includegraphics[width=8.5cm]{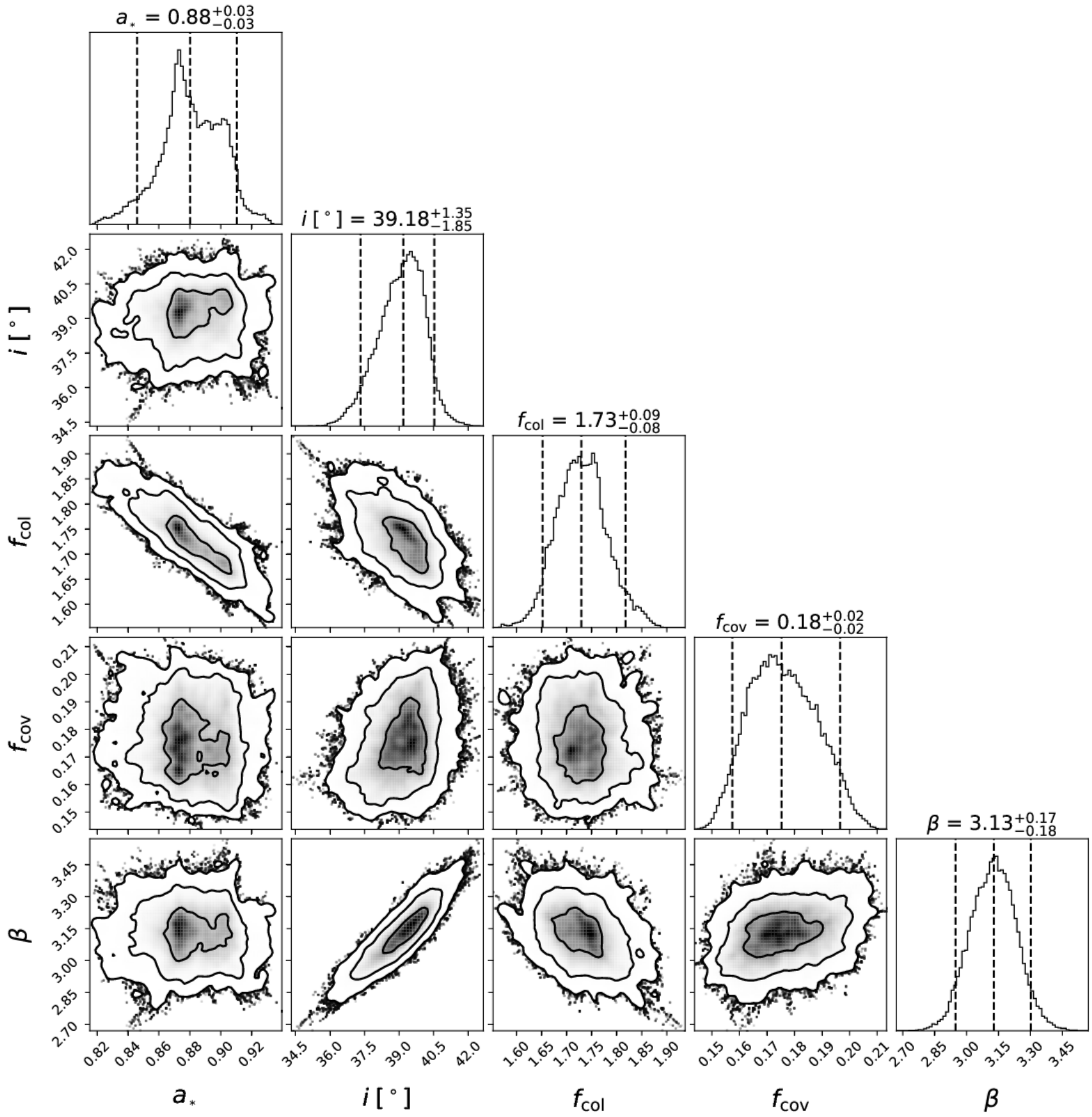}
\includegraphics[width=8.5cm]{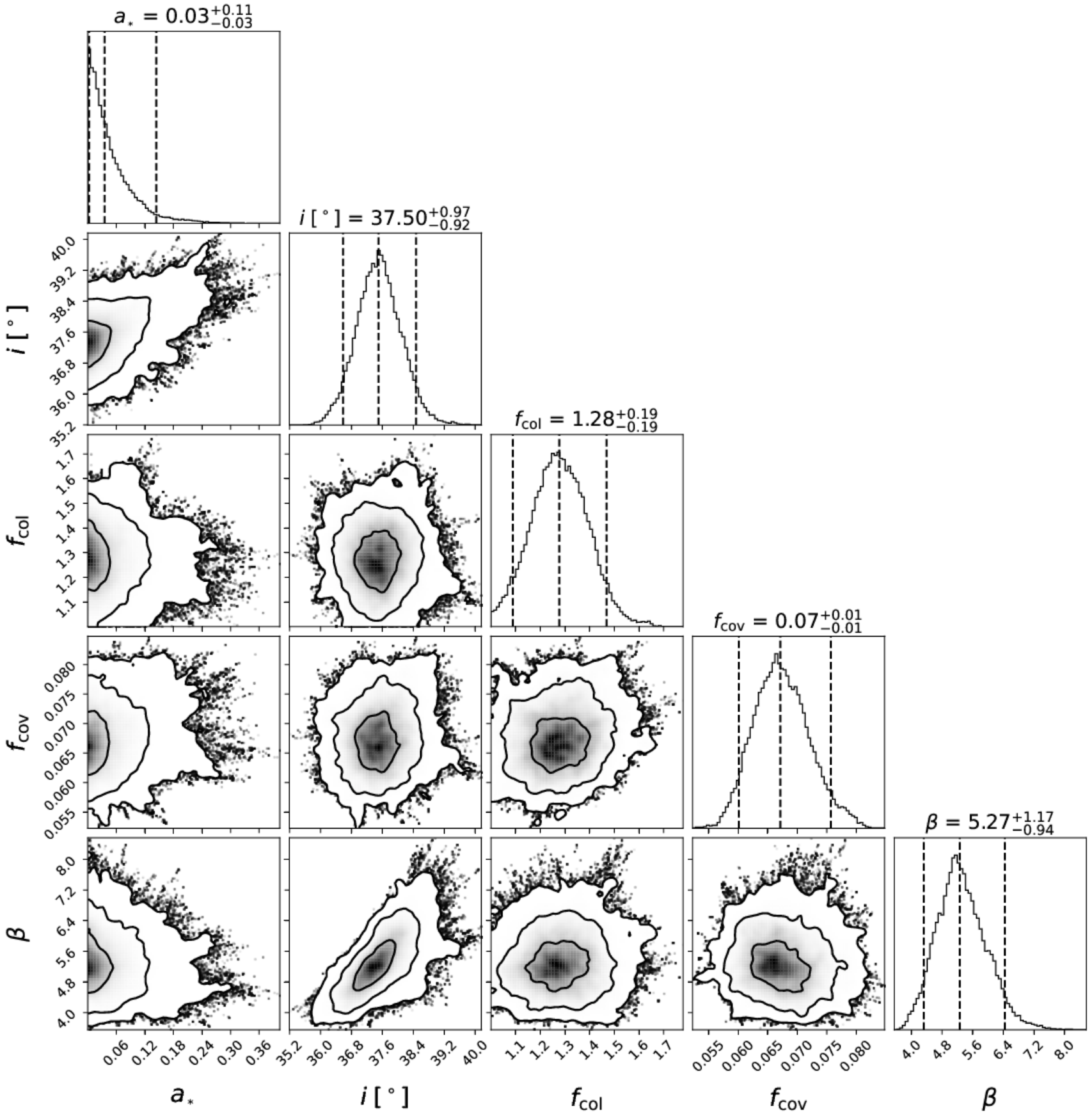}}
\caption{Correlations between selected parameters obtained by MCMC for Model 1 (left) and Model 4 (right). We show the median values by the middle dashed lines in the distribution panels and the surrounding dashed lines correspond to the 90\% uncertainty. The corresponding numerical values are given above the posterior distributions (and may be different from those in Table \ref{fits}, which give the values based on $\chi^2$).
}\label{mcmc}
\end{figure*}

We then consider a sequence of models using {\tt kerrbb} and different modelling of the high-energy tail and reflection. We begin with the approach of \citet{Tomsick14} and \citet{Walton16}. Namely, the high-energy tail is modeled by {\tt simplcut}, and the reflection features, by {\tt relxill} with a Gaussian smoothing, with the photon index and the e-folding energy tied to those of {\tt simplcut}. The Gaussian smoothing is introduced to account for the thermal broadening of the reflection from the disk with the relatively high maximum temperature of $\approx$0.4\,keV (as found by replacing {\tt kerrbb} by {\tt diskbb}; \citealt{Mitsuda84}), following \citet{Tomsick14} and \citet{Walton16}.

We also tie the spin parameters of {\tt kerrbb} (with $f_{\rm col}=1.7$) and {\tt relxill}. In order to account for distant reflection, we include a narrow Gaussian line centered at an energy allowed to be in the 6.4--7.0\,keV. However, the model can reproduce the data only very poorly, with $\chi_\nu^2=1977/612$ allowing a free inclination angle. The main reason for the poor fit is that the e-folded power law of {\tt simplcut} describes the relatively complex high-energy tail very poorly.

\subsection{High-energy tail as hybrid Comptonization + reflection}
\label{hybrid}

We next use {\tt ceqpair}. We tie the source size, $R$, and the compactness in the disk blackbody photons, $\ell_{\rm s}$, to the spin parameter. Namely, we assume the characteristic size of the disk is given by the half-power radius (dividing the regions equally contributing to the dissipation), for which we use the approximation of \citet{Fabian14},
\begin{equation}
R\approx \left[5+28(1-a_*)\right]R_{\rm g},
\label{size}
\end{equation}
where $R_{\rm g}\equiv GM/c^2$ is the gravitational radius, $\approx 3.1\times 10^6$\,cm for $M=21.2\msun$. The unabsorbed flux in the disk component is $F_{\rm disk}\approx 1\times 10^{-7}$\,erg\,cm$^{-2}$\,s$^{-1}$ (see below). At $D=2.2$\,kpc, it corresponds to the luminosity of one side of the disk of $L_{\rm disk,1/2}=\pi D^2 F_{\rm disk}/\cos i\approx 1.4\cos i^{-1} 10^{37}$\,erg\,s$^{-1}$. This corresponds to
\begin{equation}
\ell_{\rm s}\approx \frac{125}{[5+28(1-a_*)]\cos i},
\label{ls}
\end{equation}
which is in a range of $\approx$4--33 at $i=27\fdg 5$--$40\degr$ for $a_*=0$--1. For hybrid Comptonization, we assume $\gamma_{\rm min}=1.3$. We note that the amplitude of the non-thermal tail in {\tt eqpair} is determined by the $\ell_{\rm nt}/\ell_{\rm h}$ and it is not linked to $\gamma_{\rm min}$. 

We fit {\tt ceqpair(kerrbb)} together with relativistically broadened reflection of the scattered emission only (using {\tt xilconv} and {\tt relconv}) and a narrow Gaussian (accounting for distant reflection). We tie the spin parameters and the inclinations of {\tt kerrbb} and {\tt relconv}. Furthermore, we allow both the inclination and the color correction, $f_{\rm col}$, to be free parameters (the theoretical value  of the latter is uncertain, as discussed in Section \ref{intro}). We find an excellent fit, with $\chi_\nu^2=741/607$ and the parameters given in Table \ref{fits} (Model 1). The spectrum with its components are shown in Figure \ref{models}(a). In particular, it shows that a non-thermal component of the hybrid electron distribution is required, as illustrated by the plot (brown curve) of the spectral component due to scattering by thermal electrons only being well below the total scattered spectrum. We see no strong residuals at $E\gtrsim 2$\,keV, while those present at lower energies appear to be due to residual calibration inaccuracies of \nicer and an insufficient accuracy of the treatment of absorption.

The spin is $a_*= 0.87^{+0.04}_{-0.03}$, and $f_{\rm col}= 1.74^{+0.07}_{-0.09}$ includes the canonical value. The best-fit inclination of the disk and the reflection is $i= 39\degr^{+1}_{-1}$. The non-thermal injection is hard, $\Gamma_{\rm inj}= 0.2^{+0.4}_{-0.2}$, and the maximum electron Lorentz factor is low, $\gamma_{\rm max}= 6.5^{+0.4}_{-0.4}$. (We have checked that $\Gamma_{\rm inj}$ and $\gamma_{\rm max}$ remain almost unchanged when the \integral data are not included in the fit.) The fractional reflection is close to unity, ${\cal{R}}= 0.9^{+0.1}_{-0.1}$, as expected for a static corona. The scattered fraction is low, $f_{\rm sc}= 0.18^{+0.03}_{-0.04}$, i.e., the corona is patchy, which is compatible with ${\cal{R}}\sim 1$. The Fe abundance of $Z_{\rm Fe}=5.2^{+0.8}_{-0.5}$ is unrealistically high, which appears to be an artefact of using the model with the low $n=10^{15}$\,cm$^{-3}$ \citep{Garcia18n}. However, {\tt xilconv} is available for this density only. We have also tested the case with additional Gaussian broadening, but this did not improve the fit, and the best-fit Gaussian width was null.

The unabsorbed bolometric flux of this model (normalized to \nustar A) is $F_{\rm bol}\approx 1.09\times 10^{-7}$\,erg\,cm$^{-2}$\,s$^{-1}$, that in the scattered and reflected components is $F_{\rm corona}\approx 1.4\times 10^{-8}$\,erg\,cm$^{-2}$\,s$^{-1}$ and that in the disk is $F_{\rm disk}\approx 9.5\times 10^{-8}$\,erg\,cm$^{-2}$\,s$^{-1}$. Thus, the disk emits $\approx 88\%$ of the total flux, and the state is indeed disk-dominated. 

As in the model in Section \ref{powerlaw}, our physical models include a narrow Fe K$\alpha$ emission line. The line is relatively weak, with the equivalent width with respect to the total continuum of $\approx$13\,eV. As discussed in \citet{Walton16}, this component can be emitted either by the stellar wind or the surface of the donor. In the latter case, the line would be accompanied by a (nearly rest-frame) reflection continuum. In order to test this possibility, we have replaced the line by {\tt xilconv(ceqpair(kerrbb))}. This has led to virtually no changes in the fitted parameters, in particular to the spin and inclination, but it significantly increased the $\chi^2$. Thus, we keep our modelling with the narrow-line component.

We then consider a model with the inclination fixed at $i=27\fdg 5$, which we denote as Model 2. This yields $\chi_\nu^2=760/608$, i.e., $\Delta \chi^2=+19$ with respect to the model with free $i$. Using the F-test, the fit improvement at a free $i$ has the probability of being by chance of $\approx 10^{-4}$. The model yields $a_*= 0.90^{+0.02}_{-0.01}$ and a relatively large color correction $f_{\rm col}\approx 1.95^{+0.01}_{-0.03}$, as well as ${\cal{R}}= 1.2^{+0.1}_{-0.1}$.

In the next step, we also fix the color correction at the canonical value, $f_{\rm col}=1.7$, denoting this model as Model 3. This leads to a slightly larger $\chi_\nu^2=765/609$, but it significantly increases the spin, to $a_*= 0.986^{+0.002}_{-0.001}$.

We then added a warm skin layer, simulated using {\tt thcomp}, to Model 1, and we denote it as Model 4. The model yields $\chi_\nu^2=705/606$, i.e., a significantly better fit than that of Model 1, with a very low, $\approx\! 3\times 10^{-7}$, probability of the fit improvement being by chance. It does have a strong effect on the fitted spin parameter, which is now very low, $a_*= 0^{+0.07}$. The spectrum and its components are shown in Figure \ref{models}(b), and the parameters are given in Table \ref{fits}. The disk in this model has a low $f_{\rm col}= 1.26^{+0.17}_{-0.08}$. The usual value of $\sim$1.7 is caused by scattering in a top part of the disk, which decreases the efficiency of thermalization. But now the disk is covered by the optically thick and dissipating layer, diffusively increasing both the photon density and the temperature at the disk--layer boundary. This, in turn, may decrease $f_{\rm col}$. We have also fitted a model with the fixed $f_{\rm col}=1.7$, but found the parameters almost unchanged. We have also found that the low spin obtained in this model does not change when we assume the fixed $i=27\fdg5$. 

The unabsorbed bolometric flux of Model 4 is $F_{\rm bol}\approx 1.23\times 10^{-7}$\,erg\,cm$^{-2}$\,s$^{-1}$, that emitted by the disk and the warm skin $F_{\rm disk,skin}\approx 1.09\times 10^{-7}$\,erg\,cm$^{-2}$\,s$^{-1}$ and that in the underlying disk alone (underneath the warm skin) is $F_{\rm disk}\approx 5.4\times 10^{-8}$\,erg\,cm$^{-2}$\,s$^{-1}$. This implies that the disk and the warm skin contribute to the observed flux about equally. 

We then show correlations between selected parameters as well as their distributions for Models 1 and 4, see Figure \ref{mcmc}. For that, we used the Markov Chain Monte Carlo (MCMC) method \citep{Foreman-Mackey13}, as implemented in {\sc xspec}. We assume wide normal priors centered on the best-fit parameters. For Model 1, we see significant anticorrelations of $f_{\rm col}$ with $a_*$ (see \citealt{Salvesen21}), $i$, and $\beta$. We also see a strong positive correlation of $i$ with $\beta$. For Model 4, se see only two positive correlations, of $i$ with $a_*$ and with $\beta$.

We have also tried to fit the data with the hybrid-Compton model {\tt comppsc}. However, we have been unable to obtain a good fit. That model is accurate for the assumed shape of the electron distribution, which is a Maxwellian truncated at $\gamma_{\rm min}$ and a power law above $\gamma_{\rm min}$. On the other hand, the high-energy tail in {\tt ceqpair} is not a pure power law, but it is given by the solution following from the injected power-law distribution and the energy losses taking into account all the relevant processes for given $\ell_{\rm s}$, $\ell_{\rm h}$ and $\ell_{\rm nth}$. Apparently, theoretical spectra with that set of assumptions provide a good fit, while those with the assumptions of {\tt comppsc} do not.

\section{Discussion}
\label{discussion}

\subsection{The spin model dependence}
\label{spin}

Our Model 3, with the standard disk model with $f_{\rm col}=1.7$ and $i=27\fdg 5$ yields a very high spin, $a_*= 0.986^{+0.002}_{-0.001}$, see Table \ref{fits}. We can compare it to the result of \citet{Zhao21_CygX1} and \citet{Miller-Jones21}, who obtained $a_*>0.9985$ at the $3\sigma$ level, obtained by spectral fitting the disk continuum \citep{Zhao21_CygX1, Miller-Jones21}. We note that that result implies the spin even above the maximum value of 0.998 due to the radiation swallowed by the BH \citep{Thorne74}. The difference with respect to \citet{Zhao21_CygX1} and \citet{Miller-Jones21} can be accounted for by their using the {\tt kerrbb2} model and our fit also using the constraint from relativistic broadening of the reflection. Thus, we basically confirm their results under their assumptions. 

However, as discussed in Section \ref{intro}, the {\tt kerrbb2} spectral model uses the calculations of the local disk spectra of \citet{Davis05} and \citet{Davis06} for the viscosity parameter \citep{SS73} of $\alpha=0.1$. Thus, it is based on the canonical model of radiative transfer with relatively weak magnetic field generated by the magneto-rotational instability dynamo (as described, e.g., in \citealt{Davis19}). That disk model is unable to reproduce a large number of astrophysical phenomena (see Section \ref{intro}). To solve this problem, models with support by large-scale magnetic fields have been proposed, e.g., the so-called magnetically-elevated disk \citep{Begelman07}. The models can be approximately described by the color correction of $f_{\rm col}\gtrsim 1.7$ \citep{Salvesen21}.
 
Then, relaxing the assumption of the constant $f_{\rm col}= 1.7$ leads to $a_*= 0.90^{+0.02}_{-0.01}$, which allows for significantly lower values than $a_*= 0.986^{+0.002}_{-0.001}$ obtained before, at a higher $f_{\rm col}= 1.95_{-0.01}^{+0.03}$ , see our Model 2. Furthermore, the binary inclination of $i_{\rm b}=27\fdg 5$ might be different from the inclination of the inner disk. Most of the X-ray spectral fits of Cyg X-1 obtained larger inclinations (e.g., \citealt{Tomsick14, Walton16}), as well as the relatively large X-ray polarization in the hard spectral state \citep{Krawczynski22} implied at least $i\gtrsim \!40\degr$. Thus, we allowed a free inclination in our Model 1, obtaining a significantly better fit with $i= 39\degr^{+1}_{-1}$, $a_*= 0.87^{+0.04}_{-0.03}$ and $f_{\rm col}= 1.74^{+0.07}_{-0.09}$. We see that this range of $a_*$ is actually relatively similar to that of Model 2, implying that allowing for a free $i$ does not significantly change the fitted spin, though if both $i=27\fdg 5$ and $f_{\rm col}= 1.7$ are imposed, the allowed range of $a_*$ is narrow and close to unity (Model 3). Our result of $a_*\sim 0.9$ is also similar to many of those obtained by the reflection spectroscopy by \citet{Tomsick14}, \citet{Walton16} and \citet{Steiner24}. Concluding the above discussion, the spin parameter of Cyg X-1 is likely to be $a_*\sim 0.9$, as based on our current understanding of standard optically-thick accretion disks but allowing for the presence of large scale magnetic fields. 

On the other hand, we obtain a radically different result when considering the non-standard disk model motivated by the successful modelling of soft X-ray excesses in AGNs by a dissipative optically-thick Comptonizing layer covering the disk (e.g., \citealt{Petrucci20}), see discussion in Section \ref{intro}. Such a model leads to a highly significant fit improvement with respect to the standard model, $\Delta\chi^2\approx -36$ for adding two free parameters. This model yields very low spins, $a_*= 0^{+0.07}$. The low spin in the model of the relativistic broadening is compensated by a larger value of the irradiation index, with $\beta$ increasing from $\sim$2--3 to $\sim$5, meaning that the reflection is now much more concentrated to the center. The powers dissipated in the disk and the warm layer are close to each other, see Section \ref{hybrid}. A similar result was obtained for LMC X-1 \citep{Zdziarski24a}.  

A similar model was earlier applied to \suzaku data for Cyg X-1 by \citet{Belczynski23} and \citet{Zdziarski24a}, and to \nicer and \nustar data for LMC X-1 by \citet{Zdziarski24a}, where relatively similar results were obtained. We stress that our present data are of much higher statistical quality than those data for both Cyg X-1 and LMC X-1. Still, even with the present data and based on our current understanding of accretion disks, we are unable to unambiguously prove this model to correspond to the disks in the soft state of BH XRBs.

We should consider caveats related to the specific warm skin model we used. Its form is {\tt thcomp(kerrbb)}, where {\tt thcomp} is a thermal Comptonization model in the spherical geometry with the sources of the seed photons concentrated towards the center \citep{Z20_thcomp}. On the other hand, the actual geometry is of a flat disk covered by a Comptonizing dissipative layer. This leads to a diffusive enhancement of the photon density around the disk/layer boundary. This enhancement corresponds to the flux of the disk still reaching the top of the layer in spite of strong backscattering due to its large $\tau_{\rm T}$. It increases the temperature at the boundary, which effect is not included in the present model. Further tests and improvements of the warm skin model are desirable.

In our approach, we use the method based on modeling of both the disk continuum and the relativistic broadening of the reflection spectrum, with the spins for both parts of the model required to be equal. This approach was earlier used in some models of \citet{Tomsick14} and \citet{Zdziarski24a}, and by \citet{Parker16}. We stress that this allows us to better constrain the fit parameters, in particular the spin, and we consider it to be the preferred method of spin determination. However, it requires the presence of a significant coronal emission plus reflection beyond the disk spectrum. 

\subsection{Jet models and spin}
\label{jet}

BH XRBs show radio jets of two distinctly different kinds \citep{FBG04}. In the hard state, we observe steady compact jets. During outbursts of low-mass BH XRBs, we sometimes observe ejections of discrete jets during hard-to-soft state transitions. There are two main jet models for accreting BHs. In both, the jet formation requires a poloidal magnetic field. In one, the jet is powered by differential rotation of the accretion disk, mediated by magnetic field \citep{BP82}. In the other, the magnetic field is attached to the BH, and the jet is powered by extraction of the BH spin \citep{BZ77}. The latter yields the highest jet power of any models studied so far \citep{Davis20}. Specifically \citep{Tchekhovskoy15},
\begin{equation}
P_{\rm j}\approx \frac{0.05 c}{16 \pi r_{\rm H}^2}\Phi^2 a_*^2,
\label{PBZ}
\end{equation} 
where $\Phi$ is the magnetic flux threading the BH and $r_{\rm H}$ is the horizon radius. The magnetic pressure, proportional to $\Phi^2$, is limited by the ram pressure of the accreting matter, which is, in turn, proportional to the mass accretion rate, $\dot M_{\rm accr}$. Its maximum corresponds to so-called `magnetically arrested disk' (MAD; \citealt{BK74, Narayan03}), when accretion becomes choked \citep{McKinney12}. GRMHD simulations \citep{Tchekhovskoy11} yield a maximum value of $\Phi$ for geometrically thick MAD disks being weakly dependent of $a_*$. Then, a simple overall approximation is obtained,
\begin{equation}
P_{\rm j,max}\approx 1.3 \dot M_{\rm accr}c^2 a_*^2.
\label{Pj}
\end{equation} 
Thus, $a_*\sim 1$ is needed to achieve the maximum jet power. Equation (\ref{Pj}) with $a_*=1$ is often used to check whether the jet power estimated from observations agrees with the limit set by accretion. 

However, we find that values of $a_*\ll 1$ (following from our models with warm skins) for the three known high-mass BH XRBs would not in conflict with the jet observations. LMC X-3 and M33 X-7 have no jets. Cyg X-1 has a prominent compact jet, but its estimated power is consistent with being $\ll \dot M_{\rm accr}c^2$ \citep{Heinz06, Z_Egron22}.

On the other hand, transient jets during hard-to-soft state transitions have been observed from about 15 low-mass BH XRBs. In some cases, the ejecta have been observed up to a pc scale (e.g., \citealt{Carotenuto21}). Modelling of their propagation through the surrounding cavities and the ISM requires, at least in some cases, their power to be $\sim \dot M_{\rm accr}c^2$ (e.g., \citealt{Zdziarski23a, Carotenuto24}). Thus, at least some of low-mass BH XRBs require $a_*\sim 1$. 

\subsection{Spin-up by accretion}
\label{spinup}

In this work, we have shown that either $a_*\sim 0.9$ or $a_*\ll 1$ are possible, depending on the unknown aspects of the accretion disk physics. The former corresponds to our Models 1 and 2, with $a_*\equiv a_1\sim 0.9$. If this is the case, that spin can be either natal or acquired by accretion, with strong arguments against the former, see Section \ref{intro}. In the latter case, the BH needs to be spun up by accretion from the donor. For stellar models with efficient angular momentum transfer, the natal $a_*\equiv a_0\sim 0.1$. \citet{Bardeen70} derived the final spin due to accretion as a function of the final-to-initial BH mass ratio, $M/M_0$,
\begin{equation}
a_1=\frac{r_{\rm ISCO}(a_0)^{\frac{1}{2}}}{3 M/M_0}\left\{4-\left[\frac{3 r_{\rm ISCO}(a_0)}{(M/M_0)^2}-2\right]^{\frac{1}{2}}\right\},
\end{equation}
where $r_{\rm ISCO}$ is the ISCO radius \citep{Bardeen72} in units of $R_{\rm g}$. For $a_0=0.1$, $a_1=0.9$, $(M-M_0)/M_0\approx 0.56$. However, the accreted mass is $\Delta M>M-M_0$ due to the released binding energy. \citet{Bardeen70} obtained
\begin{align}
\frac{M}{M_0}=&\left[\frac{3 r_{\rm ISCO}(a_0)}{2}-1\right]^{\frac{1}{2}}\sin\left\{\left[\frac{2}{3 r_{\rm ISCO}(a_0)}\right]^{\frac{1}{2}}\frac{\Delta M}{M_0}\right\}+\nonumber\\
&\cos\left\{\left[\frac{2}{3 r_{\rm ISCO}(a_0)}\right]^{\frac{1}{2}}\frac{\Delta M}{M_0}\right\},
\end{align}
from which we obtain $\Delta M/M_0\approx 0.63$. Then, a fraction of $\approx$0.40 ($\approx 8.5\msun$ for the BH mass of $21.2\msun$) of the present mass needed to be accreted. 

If we neglect the time dependence of the accretion efficiency, $\eta$,
the Eddington-limited accretion increases the BH mass exponentially at the Eddington time,
\begin{equation}
t_{\rm E}=\frac{\eta (1+X) c \sigma_{\rm T}}{8\pi (1-\eta) G m_{\rm p}},
\label{tedd}
\end{equation}
where $X$ is the H mass fraction, $\sigma_{\rm T}$ is the Thomson cross section, $G$ is the gravitational constant and $m_{\rm p}$ is the proton mass. At $\eta=0.1$, $X=0.5$, it is $\approx$38\,Myr. The time required to increase the mass from $M_0$ to $M$ is $t_{\rm E}\ln (M/M_0)\approx 17$\,Myr. This is much longer than the estimated lifetime of Cyg X-1 of $\sim$4\,Myr \citep{Miller-Jones21}. 

The Eddington-limited accretion onto the BH was proposed by \citet{SS73}, in which case the gravitational energy released in the flow in the excess of the Eddington rate is used to eject the excess mass. On the other hand, \citet{Qin22} assumed super-Eddington accretion in Cyg X-1 to be fully conservative. They performed detailed evolutionary study for such a case. The peak mass transfer rate in their model is $\dot M\approx 10^{-2}\msun/{\rm yr}\approx 6\times 10^{23}$\,g s$^{-1}$, corresponding to $\dot M c^2\approx 6\times 10^{44}$\,erg\,s$^{-1}$. They assumed the accretion is fully conservative. 

Efficient super-Eddington accretion onto a BH can be facilitated by trapping of photons radiated by the flow, which are then advected to the BH \citep{Katz77, Abramowicz88}. In that model, the advection happens within a trapping radius, at which the photon diffusion and accretion time scales are equal. Then, all of the mass enters the BH. In reality, there are both advection and outflow, see \citet{Poutanen07} and references therein. In their model, about a half of the mass of the accretion flow enters the BH, which appears sufficient to spin up the BH. 

On the other hand, \citet{Qin22} claimed that conservative super-Eddington accretion can occur due to neutrinos carrying away the excess accretion energy. Here, we test this assertion. We have adopted parameters of super-Eddington accretion based on \citet{Poutanen07}. The spherization radius at the above $\dot M$ is $\approx 3\times 10^{12}$\,cm, and the mass within it is $\approx 10^{-5}\msun$.  We approximated it by a quasi-spherical, low angular momentum, supersonic accretion flow, which can be partly cooled by neutrinos produced mainly by electron-positron pair annihilation (\citealt{Janiuk19}; using the rates of \citealt{Rosswog03}). Neutronization of matter is negligible, due to low density of the matter, hence the resulting number densities of e$^-$, e$^+$, and heavy-lepton neutrinos are nearly equal. We have found the neutrino luminosity is low, $\lesssim\! 10^{40}$\,erg\,s$^{-1}$. This is much too low to be able to affect the system dynamics at the above $\dot M c^2$. Thus, we do not confirm the statement of \citet{Qin22} that neutrino cooling allows super-Eddington accretion to operate.

Still, as we discussed above, relatively efficient accretion can occur instead due to photon trapping acting together with outflows, as in the model of \citet{Poutanen07}. Thus, the spin of $a_*\sim 0.9$ could have been acquired by super-Eddington accretion, provided such an evolutionary path is confirmed by calculations of the stellar evolution.

\section{Conclusions}

Our main conclusions are as follows.

We have performed a detailed study of the highly accurate broad-band X-ray data on Cyg X-1 obtained in the soft spectral state by simultaneous NICER and \nustar observations and a contemporaneous \integral observation. We found the shape of the high-energy tail is relatively complex. It was very poorly described by a power law with reflection. The best model we have found is that of hybrid Comptonization, which we modeled by a convolution version of {\tt eqpair}, with the seed photons from a relativistic disk, modelled by {\tt kerrbb}. In addition, the spectrum showed relativistically broadened reflection features at the strength approximately corresponding to a slab geometry. We tied the spin parameters of the disk and relativistic broadening models, thus combining the disk continuum and the reflection spectroscopy methods of spin determination. 

We have found the measured BH spin parameter of Cyg X-1 to be strongly dependent on the way the disk is modeled. For the standard accretion disk model \citep{NT73, Davis05, Davis06}, in which the magnetic fields result from the magneto-rotational dynamo and for the binary inclination inferred from optical studies, our measurements basically confirm the previous results of $a_*\gtrsim 0.99$. 

However, the standard model fails to explain a number of major astrophysical phenomena, as discussed in Section \ref{intro}. Then, disks can be partially supported by magnetic pressure (e.g., \citealt{Begelman07, Begelman17}), which can increase the color correction. Allowing for a free $f_{\rm col}$ leads to $a_*= 0.90^{+0.02}_{-0.01}$ and $f_{\rm col}= 1.95^{+0.01}_{-0.03}$. Furthermore, if the disk inclination is allowed to be different from $i_{\rm b}$, e.g., due to a warping, $a_*= 0.87^{+0.04}_{-0.03}$ and $f_{\rm col}= 1.74^{+0.07}_{-0.09}$, $i= 39\degr^{+1}_{-1}$, with the fit significantly improving.

In addition, we consider the case of the dissipation occurring also in a warm surface layer, as motivated by successful modeling of soft X-ray excesses in AGNs in this way. Then, Comptonization in that layer leads to very low spins, $a_*= 0^{+0.07}$ with a very low chance probability for the fit improvement, of $3\times 10^{-7}$. This result can resolve the tension between the low BH spins inferred from analyses of merger events detected in gravitational waves and the prevalence of high spins estimated by spectral fitting of BH XRBs. 

\section*{Acknowledgements}
We thank Chris Belczy{\'n}ski, Tomek Bulik, Shane Davis, P.O.\ Petrucci and Jack Steiner for valuable discussions, and the referee for valuable comments. We acknowledge support from the Polish National Science Center under the grants 2019/35/B/ST9/03944, 2023/48/Q/ST9/00138 and 2023/50/A/ST9/00527, and from the Copernicus Academy under the grant CBMK/01/24. MS acknowledges support from University of {\L}{\'o}d{\'z} IDUB grant, decision No.\ 59/2021.

\bibliographystyle{aasjournal}
\bibliography{../../allbib} 

\end{document}